\documentclass[aps,preprintnumbers,eqsecnum,amsmath,amssymb,nofootinbib]{revtex4}
\usepackage{graphicx}
\usepackage[english]{babel}
\usepackage{url}
\usepackage{graphicx}
\usepackage{xcolor}
\usepackage{amsmath}
\usepackage{amssymb}
\usepackage{slashed}
\usepackage{multirow}
\begin{document}
\title{{\boldmath Probing dark-matter scenarios of Belle II excess in {\boldmath $B^+\to K^+M_X$}\\ with {\boldmath $K^+\to \pi^+ M_X$} decays}}
\author{Alexander Berezhnoy$^{a}$, Wolfgang Lucha$^{b}$, and Dmitri Melikhov$^{a,c,d}$}
\affiliation{
$^a$D.~V.~Skobeltsyn Institute of Nuclear Physics, M.~V.~Lomonosov Moscow State University, 119991, Moscow, Russia\\
$^b$Marietta Blau Institute (MBI) for Particle Physics, Dominikanerbastei 16, A-1010 Vienna, Austria\\
$^c$Joint Institute for Nuclear Research, 141980 Dubna, Russia\\
$^d$Faculty of Physics, University of Vienna, Boltzmanngasse 5, A-1090 Vienna, Austria}
\date{\today}
\begin{abstract}
The hypothesis of top-philic dark matter as the origin of the Belle II excess events in $B\to K M_X$ decays has direct impact on $K\to \pi M_X$ decays: Both of these decays proceed via loop diagrams~with a top quark emitting some mediator field $R$ and the subsequent decay of $R$ into a pair of dark-matter fermions $\bar\chi\chi$. The Belle II results on $B\to K M_X$ allowed us to extract the parameters of the mediator $R$ with reasonable accuracy, whereas the sensitivity to the fermion mass $m_\chi$ proved to be rather low. In contrast to $B\to K M_X$ decays, the $K\to \pi M_X$ decays exhibit a strong sensitivity to $m_\chi$; therefore, if---upon reducing theoretical and experimental uncertainties---the tension between standard-model predictions for the $K\to\pi M_X$ decays and the most recent NA62 measurements becomes pronounced the latter decays may either enable the extraction of the value of $m_\chi$ or provide a lower limit on~$m_\chi$.
\end{abstract}
\maketitle

\section{\label{Sect:1}Introduction}
In the Standard Model (SM), the missing-energy decay $B^+\to K^+ M_X$ is effected by the decays $B^+\to K^+\bar\nu\nu$ with a pair of neutrino $\nu$ and antineutrino $\bar\nu$ in the final state. This flavour-changing neutral-current (FCNC) induced~process is, in the SM, forbidden at tree level and proceeds via loops (penguins and boxes). For the resulting branching ratios, rather accurate theoretical predictions have been reported \cite{damir2023epjc,hpqcd2023,Allwicher:2023xba}:
\begin{eqnarray}
 \label{damir}
Br(B^+\to K^+\bar\nu\nu)=(4.44\pm 0.30) \times 10^{-6}, \qquad
Br(B^+\to K^{*+}\bar\nu\nu)=(9.8\pm 1.4) \times 10^{-6}.
\end{eqnarray}
In 2023, the Belle II collaboration reported \cite{Belle-II:2023esi} a surprising observation significantly exceeding the SM expectation,~viz.,
\begin{eqnarray}
Br(B^+\to K^+\bar\nu\nu)&=&(2.3\pm 0.7) \times 10^{-5}=(5.4\pm 1.5)Br(B^+\to K^+\bar\nu\nu)_{\rm SM},
\end{eqnarray}
as well as, in addition, the measurement of the differential distribution for the decay $B\to K M_X$. If confirmed,~these findings require the contribution of new physics beyond the SM. One of many new-physics scenarios for the $B\to K\bar\nu\nu$ enhancement is provided by the interpretation of the Belle II excess events as $B$ decays into invisible dark-matter~(DM) particles \cite{Abdughani:2023dlr,Berezhnoy:2023rxx,datta,Calibbi:2025rpx,He:2024iju,Hou:2024vyw,Gabrielli:2024wys,Ho:2024cwk,
 He:2025jfc,Kolay:2025jip,blm2025,Aliev:2025hyp,Ding:2025eqq,He:2025zfy,blm2026,universe,Liu:2025lbw,Kim:2025zaf,Mescia:2026xju, Bolton:2025lnb,Jahedi:2026dzy,Gartner:2026clx,Ho:2026kqp,Berezhnoy:2026yii}. Various specific scenarios have been considered, involving, for instance, a dark photon or decays~into a DM fermion, $\chi$, and antifermion, $\bar\chi$, via some mediator field, in the following generically labelled $R$. Taking into account the additional DM contribution, the $B\to K M_X$ decay width becomes
\begin{eqnarray}
\Gamma(B\to K M_X)=\Gamma(B\to K \bar\nu\nu)_{\rm SM}+\Gamma(B\xrightarrow{R}K \bar\chi\chi).
\end{eqnarray}

In the latter context, in our respective previous analyses \cite{Berezhnoy:2023rxx,blm2025,blm2026,universe,Berezhnoy:2026yii} we studied $B\to K^{(*)} M_X$ decays within two specific top-philic DM scenarios defined by assuming the mediator field $R$ coupling both to the top quark $t$ and~to~a DM fermion-antifermion pair $\chi\bar\chi$, for either a scalar mediator field $R_s$ (labeled S-scenario) or a vector~mediator field~$R_v$ (denoted V-scenario). For both scenarios, a good description of the Belle II excess data could be demonstrated and~all relevant DM parameters (that is, the mediator mass $M_R$, the mediator decay width, and the couplings of the mediator to the top quark, $g_{Rtt}$, and to the DM fermion, $g_{R\chi\chi}$) have been extracted.

Noteworthy, within top-philic scenarios the same mechanism contributes to any FCNC $b\to s$ and $s\to d$ transition, in particular, to the missing-energy decay $K^+\to\pi^+ M_X$, with the evident consequence for the latter's decay width~that
\begin{eqnarray}
\Gamma(K\to \pi M_X)=\Gamma(K\to \pi \bar\nu\nu)_{\rm SM}+\Gamma(K\xrightarrow{R}\pi \bar\chi\chi).
\end{eqnarray}
Both $K^+\to\pi^+\bar\nu\nu$ SM amplitude and corresponding differential distribution are known quite well (cf., e.g., the~classical work \cite{Buchalla:1995vs} and its recent update \cite{Brod:2021hsj}). For the additional contribution $\Gamma(K^+\xrightarrow{R}\pi^+ \bar\chi\chi)$, upon adopting the top-philic DM parameter values extracted from the successful description of the Belle II excess events in the $B\to K M_X$ decays, one readily obtains parameter-free predictions. We calculate the differential distribution of the decay $K\xrightarrow{R}\pi \bar\chi\chi$~within both S- and V-scenarios and study further constraints on the DM sector, which may be obtained from a comparison~of our theoretical predictions with the NA62 experimental results for $K^+\to \pi^+ M_X$.

\section{\label{Sect:2}Standard-Model and Dark-Matter Contributions to the $K^+\to\pi^+ M_X$ Decay}

\subsection{$K^+\to\pi^+\bar\nu\nu$ Decays in the SM}
The effective Hamiltonian $H_{\rm eff}$ describing the $s\to d\bar\nu\nu$ transition in the SM has been available for thirty years~\cite{Buchalla:1995vs}:
\begin{eqnarray}
 H_{\rm eff}(s\to d \bar\nu\nu)&=&c_L \bar s\gamma_\mu(1-\gamma_5)d\cdot \bar\nu\gamma^\mu(1-\gamma_5)\nu,\\
 c_L&=&\frac{G_F}{\sqrt{2}}\frac{\alpha_{\rm em}}{2\pi \sin^2\theta_W}\sum\limits_{l=e,\mu,\tau}
 \left(V^*_{cb}V_{cs}X^l+V^*_{tb}V_{ts}X_t\right).
\end{eqnarray}
The corresponding differential $K^+\to\pi^+\bar\nu\nu$ decay rate, with respect to the square $q^2$ of the difference of the involved meson momenta, $q^2=(p_{K^+}-p_{\pi^+})^2$, reads \cite{colangelo1996,Melikhov:1997wp}
\begin{eqnarray}
 \frac{d\Gamma(K^+\xrightarrow{\rm SM}\pi^+\bar\nu\nu)}{dq^2}=
 \frac{c_L^2}{32\pi^3M_K^3}\lambda^{3/2}(M_K^2,M_\pi^2,q^2)|f^{K\to\pi}_+(q^2)|^2,
\end{eqnarray}
where $\lambda(a,b,c)=(a-b-c)^2-4bc$. The analytic expressions for $X^l$ and $X_t$ can be found in Sec.~XI of \cite{Buchalla:1995vs}. We use the numerical values of all the involved parameters from the comparatively recent analysis \cite{Brod:2021hsj}, which yield
\begin{eqnarray}
c_L=(3.1\pm 0.14)\times 10^{-11}\mbox{ GeV$^{-2}$},
\end{eqnarray}
corresponding to
\begin{eqnarray}
Br(K^+\to \pi^+\bar\nu\nu)=(7.71\pm 0.61)\times 10^{-11}.
\end{eqnarray}
The uncertainty in this estimate arises primarily from the uncertainty in $c_L$, since the form factor $f^{K\to\pi}_+(q^2)$ is known to a very good accuracy from lattice QCD \cite{FLAG:2024oxs}:
\begin{eqnarray}
f^{K\to\pi}_+(0)=0.9698\pm 0.0017.
\end{eqnarray}
The uncertainty in the $q^2$-dependence of this form factor plays a minor role since the relevant range of $q^2$ is limited, $0<q^2<(M_K-M_\pi)^2$, and the maximal $q^2$ is much less than the mass of the lowest resonance in the $q^2$-channel, $K^*$, that is, $(M_K-M_\pi)^2\ll M_{K^*}^2$.

\subsection{$K^+\to\pi^+\bar\chi\chi$ Decays in the DM S- and V-Scenarios}
\begin{center}
\begin{figure}[h!]
 \includegraphics[width=8cm]{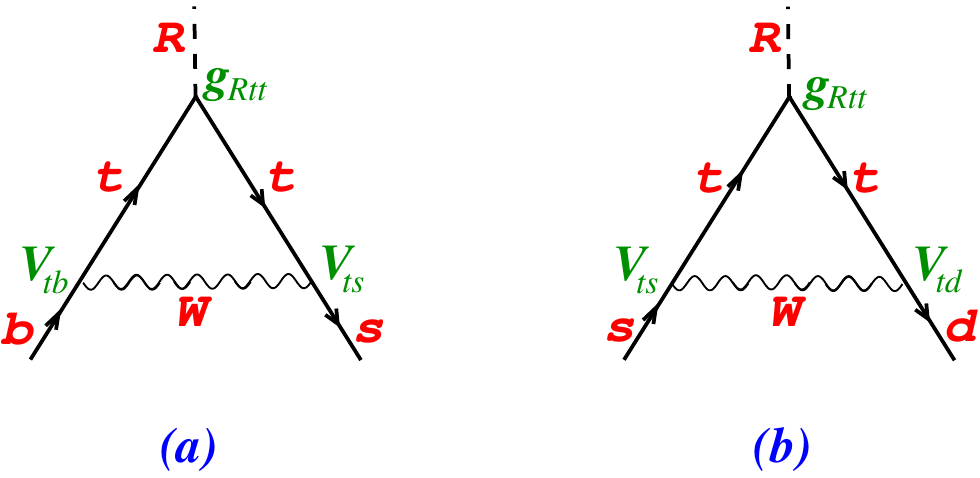}
 \caption{\label{Fig:1}Feynman graphs underlying the FCNC decays $b\to s\bar\chi\chi$ (a) and $s\to d\bar\chi\chi$ (b) in top-philic DM scenarios. The Wilson coefficients corresponding to Diagram (a) and Diagram (b) are related to each other by the simple replacements $V_{tb}V_{ts}\to V_{ts}V_{td}$ of the CKM factors and $b\to d$ of the involved quark fields.}
\end{figure}
\end{center}

For the decay $K^+\to\pi^+ M_X$, we work out implications of the conjecture that the Belle II excess events actually~arise from the decay into a pair of $\bar\chi\chi$ DM fermions that couple to the top quark via some mediator field $R$. No interaction of the mediator $R$ with other SM particles is implied. In this case, both FCNC decays $b\xrightarrow{R}s\bar\chi\chi$ and $s\xrightarrow{R}d\bar\chi\chi$~proceed via the top--$W$-boson loop depicted in Fig.~\ref{Fig:1}. Therefore, the amplitude of the $K^+\to\pi^+\bar\chi\chi$ decay is obtained from the amplitude of the $B^+\to K^+\bar\chi\chi$ decay by obvious replacements of the CKM factors, quark masses, and form factors.

\subsubsection{S-Scenario}
A rather popular model \cite{Batell:2009jf, Schmidt-Hoberg:2013hba}, referred to as $S$-scenario, involves the interaction of some DM fermions $\chi$ with the top quark $t$ by exchange of a scalar mediator field $R_s$, governed by the interaction Lagrangian
\begin{eqnarray}
 \mathcal{L_{\rm int}} = - \frac{g_{R_stt} m_t}{v} R_s\, \bar t t - g_{R_s\chi\chi} R_s\, \bar\chi\chi;
 \label{eq:lagrangian}
\end{eqnarray}
here $v\simeq 246$ GeV is the vacuum expectation value of the SM Higgs field. The emerging effective Lagrangian~encoding the FCNC vertex $b\to sR_s$ then reads \cite{Batell:2009jf, Schmidt-Hoberg:2013hba}
\begin{eqnarray}
\label{Leff}
\mathcal{L}_{s \rightarrow dR_s} = g_{sdR_s}\,R_s\, \bar d_L s_R + {\rm h.c.}, \qquad
 g_{sdR_s}=\frac{g_{R_stt} m_s}{v} \frac{3 \sqrt{2} G_F m_t^2 V^*_{td} V_{ts}}{16 \pi^2}.
\end{eqnarray}
As depicted in Fig.~\ref{Fig:1}(b), the $K^+\to \pi^+\bar\chi\chi$ transition proceeds via the mediator field $R_s$. One of the ingredients of~the amplitude of interest, describing the $K^+\to \pi^+$ transition and expressed in terms of a form factor $f_0^{K\to \pi}(q^2)$, assumes the form
\begin{eqnarray}
\label{FF}
 \langle \pi^+|\bar d_L s_R|K^+ \rangle &=&\frac{1}{2}\langle \pi^+|\bar d s|K^+\rangle=
 \frac{1}{2} \frac{M_K^2-M_\pi^2}{m_s-m_d}f_0^{K\to \pi}(q^2).
\end{eqnarray}

For the differential distribution in the $S$-scenario, upon neglecting $m_d$ compared with $m_s$, we obtain \cite{blm2026}
\begin{eqnarray}
\frac{d\Gamma(K^+\xrightarrow{\rm S} \pi^+ \bar\chi\chi)}{dq^2}&=&
\frac{\lambda^{1/2}(M_K^2,M_\pi^2,q^2)}{128 \pi^3 M_K^3}
\frac{(M_K^2-M_\pi^2)^2 |f_0^{K\to \pi}(q^2)|^2}{4m_s^2}
\frac{g_{sdR_s}^2\,g_{R_s\chi\chi}^2}{(M_{R_s}^2-q^2)^2+M_{R_s}^2\Gamma_{R_s}^2(q^2)}
 q^2 \left (1-\frac{4 m_\chi^2}{q^2}\right)^{3/2},\nonumber\\
\label{eq:dGdq2}
\end{eqnarray}
with $\Gamma_{R_s}(q^2)$ representing the $q^2$-dependent width of the mediator $R_s$ \cite{blm2025,blm2026}. This $\Gamma_{R_s}(q^2)$ width can be inferred~from the imaginary part of the fermion-loop Feynman diagram with scalar vertices (cf. \cite{gs,nachtmann}) in the form of
\begin{equation}
\label{Gamma}
\Gamma_{R_s}(q^2)= \left(\frac{q^2-4m_\chi^2}{M_{R_s}^2-4m_{\chi}^2}\right)^{\frac{3}{2}}\frac{M_{R_s}}{\sqrt{q^2}}\; \Theta(q^2-4m_\chi^2)\;
\Gamma_{R_s}^0, \qquad \Gamma_{R_s}^0=\frac{g^2_{R_s\chi\chi}}{8\pi}M_{R_s}\left(1-\frac{4m_\chi^2}{M_{R_s}^2}\right)^{\frac{3}{2}}.
\end{equation}
We emphasize that the $s$-quark mass $m_s$ in the denominator of Eq.~(\ref{eq:dGdq2}) cancels the factor $m_s$ in $g_{sdR_s}$, Eq.~(\ref{Leff}); it thus drops out from the decay rate. The above expressions involve four DM parameters: $M_{R_s}$, $m_\chi$, $g_{sdR_s}$, and $g_{R_s\chi\chi}$. Below, we assume that $M_{R_s}> 2 m_\chi$ and, furthermore, that the mediator $R_s$ decays predominantly into the $\bar\chi\chi$ pair.

\subsubsection{V-Scenario}
A widely discussed top-philic vector-mediator scenario, involving a vector mediator field $R_v^\mu$ that couples both to the top quark $t$ and to some DM fermions $\chi$, is defined by the Lagrangian
\cite{Langacker:2008yv,Cox:2015afa,Hu:2024xes}
\begin{eqnarray}
 \mathcal{L}_{\rm int}=g_{R_vtt}\,R_v^\mu\,\bar t \gamma_\mu (1+\gamma_5)t+g_{R_v\chi\chi}\,
 R_v^\mu\,\bar\chi \gamma_\mu \chi.
\end{eqnarray}
The coupling of the mediator $R_v^\mu$ to the $s\to d$ FCNC occurs by means of the penguin loop diagram shown in Fig.~\ref{Fig:1}(b). Integrating out the heavy top and $W$ fields leads to the $s\to dR_v$ effective Lagrangian \cite{Inami:1980fz}
\begin{eqnarray}
\label{LbsV}
\mathcal{L}_{s \rightarrow dR_v}=g_{sdR_v} \bar d\gamma_\mu (1-\gamma_5)s\, R_v^\mu.
\end{eqnarray}
The resulting Wilson coefficient $g_{sdR_v}\equiv \bar g_{sdR_v}(M_W)$, where $M_W$ denotes the mass of the $W$ boson in the penguin~loop of Fig.~\ref{Fig:1}(b), has the form
\begin{eqnarray}
 \label{gbsV}
 g_{sdR_v}=
 -\frac{g_{R_vtt}}{16\pi^2}\frac{G_F}{\sqrt{2}}V_{ts}V_{td}^*\,m_t^2\,F(x_t),\quad x_t=\frac{m_t^2}{M_W^2}, \quad
 F(x_t)=\frac{7-8x_t+x_t^2+2[4+(x_t-2)x_t]\log\,x_t}{(1-x_t)^2}=0.19.\nonumber\\
\end{eqnarray}
The associated differential distribution $d\Gamma(K\xrightarrow{\rm V} \pi\bar\chi\chi)/dq^2$, involving the width $\Gamma_{R_v}(q^2)$ of the mediator $R_v$, reads~\cite{blm2026}
\begin{eqnarray}
\label{dGammaDMV}
 \frac{d\Gamma(K\xrightarrow{\rm V}\pi\bar\chi\chi)}{dq^2}=
 \frac{\lambda^{3/2}(M_K^2,M_\pi^2,q^2)}{192\pi^3}\left(1+\frac{2m_\chi^2}{q^2}\right)\sqrt{1-\frac{4m_\chi^2}{q^2}}
 \frac{g_{bsR_v}^2 g_{R_v\chi\chi}^2}{(M_{R_v}^2-q^2)^2+\Gamma^2_{R_v}(q^2)M_{R_v}^2}\,|f_+^{K\to \pi}(q^2)|^2.
\end{eqnarray}
Under the assumption that $M_{R_v}>2m_\chi$, the $q^2$-dependent decay width $\Gamma_{R_v}(q^2)$ can be read off from the imaginary~part of the two-point Feynman diagram with fermions $\chi$ in the loop and vectorial vertices \cite{nachtmann}:
\begin{eqnarray}
 \label{GammaV}
 \Gamma_{R_v}(q^2)=\frac{M_{R_v}}{\sqrt{q^2}}
 \sqrt{\frac{q^2-4m_\chi^2}{M_{R_v}^2-4m_{\chi}^2}}\frac{q^2+2m_\chi^2}{M_{R_v}^2+2m_\chi^2}\Theta(q^2-4m_\chi^2)\; \Gamma_{R_v}^0,
\quad \Gamma_{R_v}^0=\frac{g^2_{R_v\chi\chi}M_{R_v}}{12\pi}\sqrt{1-\frac{4m_\chi^2}{M_{R_v}^2}} \left(1+\frac{2m_\chi^2}{M_{R_v}^2}\right).\nonumber\\
\end{eqnarray}
These expressions involve the four DM parameters $M_{R_v}$, $m_\chi$, $g_{sdR_v}$, and $g_{R_v\chi\chi}$.

\section{DM Contributions to the Observables}
For most of the DM parameters of relevance for the present analysis, their values have been extracted in our earlier studies \cite{blm2025,blm2026,universe} by fitting the Belle II observation \cite{Belle-II:2023esi} of missing-energy excess events within both S- and V-scenarios. The only DM parameter yet to be pinned down is the mediator--quark coupling $g_{sdR}$,  $R=R_s$ or $R=R_v$, respectively.

\begin{itemize}
\item For the S-scenario, the ``best'' parameter values of $B$-decay origin, corresponding to the global minimum of~$\chi^2$,~are
\begin{eqnarray}
 \label{DMS}
 M_{R_s}=2.4\pm 0.4 \mbox{ GeV},\quad\Gamma^0_{R_s}=2.9^{+1.1}_{-0.9} \mbox{ GeV},\quad m_\chi=0.42^{+0.2}_{-0.4}\mbox{ GeV},\quad g_{R_s\chi\chi}\approx 6.08.
\end{eqnarray}
In addition, the S-scenario entails a simple relation between $B$-decay relevant and $K$-decay relevant parameters:
\begin{equation}
\label{BK-S}
g_{sdR_s}=g_{bsR_s}\frac{V_{td}}{V_{tb}}\frac{m_s}{m_b}.
\end{equation}
Recalling the earlier obtained result $g_{bsR_s}\approx 5.6\times 10^{-8}$ and inserting it in Eq.~(\ref{BK-S}) yields for its $sdR_s$ counterpart
\begin{equation}
\label{Q-S}
g_{sdR_s}\approx 2.4\times 10^{-11}.
\end{equation}
\item For the V-scenario, proceeding in a manner similar to the S-scenario implies, for the $B$-decay relevant parameters,
\begin{eqnarray}
 \label{DMV}
 M_{R_v}\le 3 \mbox{ GeV},\quad\Gamma^0_{R_v}=4.0^{+2.0}_{-1.5}\mbox{ GeV},\quad m_\chi=0.6^{+0.10}_{-0.18}\mbox{ GeV},\quad g_{R_v\chi\chi}\approx 7.13.
\end{eqnarray}
In the V-scenario, the relationship between $B$-decay governing and $K$-decay governing parameters is still simpler,
\begin{equation}
g_{sdR_v}=g_{bsR_v}\frac{V_{td}}{V_{tb}},
\end{equation}
and trivially enables us to deduce the value of $g_{sdR_v}$ from the previously extracted parameter $g_{bsR_v}\approx 1.92\times 10^{-8}$:
\begin{equation}
\label{Q-V}
g_{sdR_v}\approx 1.65\times 10^{-10}.
\end{equation}
\end{itemize}

\begin{center}
 \begin{figure}[ht]
 \begin{tabular}{cc}
 \includegraphics[width=7.36cm]{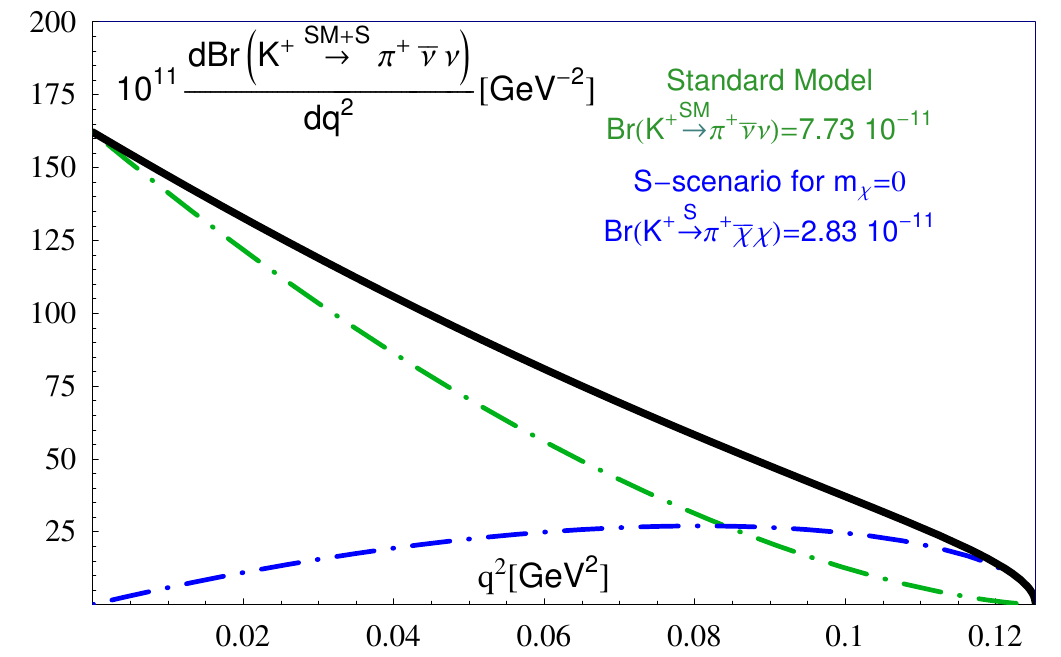}& \includegraphics[width=7.36cm]{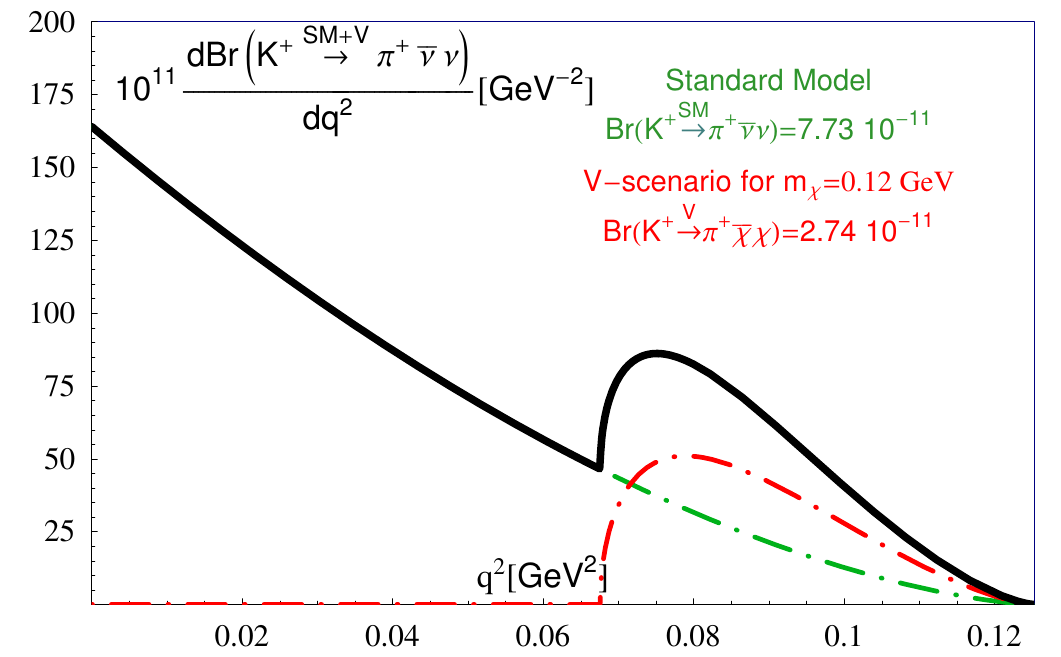}\\
 (a) & (b)\\
\includegraphics[width=7.36cm]{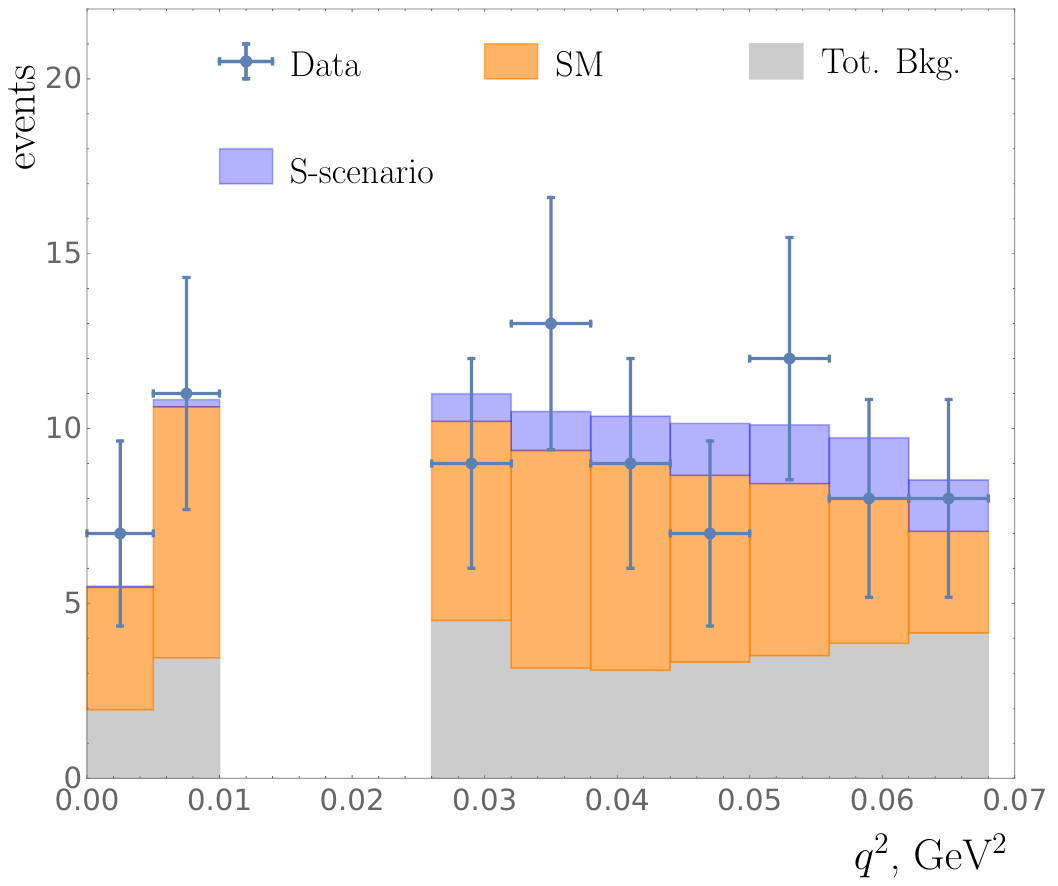}&\includegraphics[width=7.36cm]{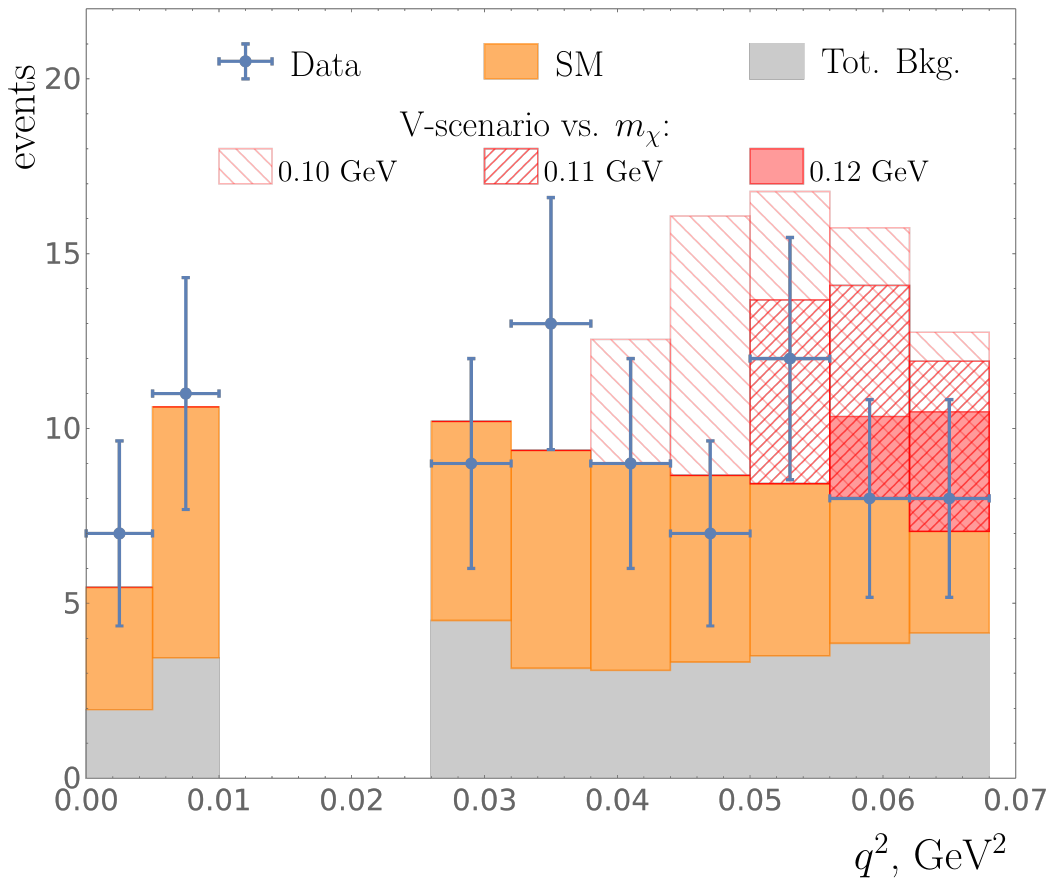}\\
 (c) & (d)
 \end{tabular}
 \caption{\label{Fig:2}Glance at missing-energy decay $K^+\to \pi^+M_X$: Confrontation of the differential distributions composed of SM and DM contributions (top row)--with dash-dotted green lines indicating the SM prediction and dash-dotted blue or red lines highlighting the additional contributions caused by the $K\to \pi \bar\chi\chi$ decay--and of the degree of agreement between the binned experimentally measured differential distribution--extracted from Fig.~8(a) of Ref.~\cite{NA62a}--with the expectation arising from combining the SM and the additional DM contributions (bottom row), for both the DM S-scenario (left column) and the DM V-scenario (right~column).}
\end{figure}
\end{center}

Figure \ref{Fig:2} summarizes the differential distributions for the missing-energy decay $K^+\to \pi^+M_X$ as obtained within~the SM and the S- and V-scenarios for the DM parameters from, respectively, Eqs.~(\ref{DMS}) and (\ref{Q-S}), or (\ref{DMV}) and (\ref{Q-V}). We separately present the theoretically predicted differential distributions in (a) the SM vs the DM S-scenario and (b) the SM vs the DM V-scenario. Moreover, the binned experimentally measured differential distribution from Ref.~\cite{NA62a} vs the SM contribution and the additional DM contribution are shown for both the S-scenario (c) and the V-scenario~(d).

Various brief comments on our insights gained, and the conclusions inferred, from the outcomes of this investigation, collectively presented in Fig.~\ref{Fig:2}, are in order:
\begin{itemize}
\item The $B^+\to K^+ M_X$ decays proved, as revealed in earlier studies \cite{blm2025,blm2026,universe}, to be not particularly sensitive~to~the DM fermion mass $m_\chi$: For the S-scenario, practically any mass from the range $0<m_\chi<0.6$ GeV is admissible. For the V-scenario, the mass constraint was more stringent: $0.2-0.3\ {\rm GeV}< m_\chi<0.7-0.8$\ {\rm GeV}. In contrast~to this, the $K^+\to \pi^+ M_X$ decays turn out to be very sensitive to the precise value of $m_\chi$ for the case of light DM fermions, satisfying $m_\chi<(M_K-M_\pi)/2$, for which the decay $K^+\to \pi^+\bar\chi\chi$ is kinematically allowed.
\item For the S-scenario, the maximal DM contribution to the $K^+\to \pi^+M_X$ differential distribution corresponds to a vanishing DM fermion mass, $m_\chi=0$. There, it amounts to some $30-40\%$ excess, more or less homogeneously smeared over the full allowed $q^2$-region. This result is well compatible with the recent NA62 measurement of the differential distributions in $K^+\to \pi^+ M_X$ \cite{NA62a,NA62b,NA62c}. Thus, the DM contribution to $K^+\to \pi^+ M_X$ in the S-scenario is well compatible with the DM parameters deduced earlier from the Belle results for the decay $B^+\to K^+ M_X$.
\item For the V-scenario, the emerging overall picture is quite different: There, a pronounced sensitivity of the excess events induced by some DM contribution in $K^+\to \pi^+ M_X$ is observed. The region $m_\chi<0.11$ GeV is practically excluded by the NA62 measurements, in excellent agreement with the constraints inferred from the Belle II~data on $B^+\to K^+ M_X$. A highly interesting case is the region $m_\chi>0.11-0.12$ GeV: There, one will note a~strong enhancement of the differential distribution in the region $q^2>0.07$ GeV$^2$. Optimists may see--or even interpret such behavior as--some indications of an enhancement of the experimental results compared to the SM predictions in this region. If this enhancement is established for greater statistics, an explanation of these findings within~the DM V-scenario will be a viable possibility.
\end{itemize}

\section{\label{Sect:5} Summary}

In conclusion, we analyzed the consequences, for the missing-energy decay $K^+\to \pi^+ M_X$, of the conjecture of some top-philic DM origin of the Belle II excess events observed for the decay $B^+\to K^+ M_X$. Within the framework of~this conjecture, the branching ratio of the decay under study receives two contributions:
\begin{eqnarray}
Br(K^+\to \pi^+M_X)=Br(K^+\to \pi^+\bar\nu\nu)_{\rm SM}+Br(K^+\xrightarrow{R} \pi^+\bar\chi\chi)_{\rm DM}.
\end{eqnarray}
Herein, the DM contribution involves precisely the same DM parameters as those determined previously \cite{blm2026} from~the analysis of the decay $B^+\to K^+ M_X$. Therefore, the decay $K^+\to \pi^+ M_X$ provides an unambiguous probe of the two top-philic scenarios employed.

The present experimental results reported by NA62 \cite{NA62a} are compatible within uncertainties with the SM predictions.
Nevertheless, an optimistic eye may tend to see some excess events in the data. If (ultimately) established at a higher statistics, such excess events might be a useful additional probe of DM (as well as of any other new physics) scenarios.

Our conclusions may be summarized as follows:
\begin{itemize}
\item[(i)]
The mediator mass $M_R$ was determined from the Belle II $B^+\to K^+ M_X$ data with reasonable accuracy, whereas the DM fermion mass $m_\chi$ extracted from these data had a larger uncertainty. The decay $K^+\to \pi^+ M_X$, however, provides a very promising possibility to probe the DM fermion mass $m_\chi$ in the region $m_\chi<(M_K-M_\pi)/2$, where the decay $K^+\to\pi^+\bar\chi\chi$ is kinematically allowed: The differential distribution of $K^+\to \pi^+\bar\chi\chi$ is sensitive~to the precise value of $m_\chi$, thus potentially providing constraints on $m_\chi$.
\item[(ii)]
For the S-scenario, the allowed range of $m_\chi$ extracted from the Belle II data on $B\to K M_X$ essentially tolerates for the mass $m_\chi$ any value in accordance with the bounds $0\le m_\chi<0.7$~GeV. The maximal DM contribution, of branching ratio $Br(K^+\xrightarrow{S} \pi^+\bar\chi\chi)=2.83\times 10^{-11}$, is expected for $m_\chi=0$. Figure~\ref{Fig:2}(a) provides the theoretical differential distributions for both the SM and the DM S-scenario, whereas Fig.~\ref{Fig:2}(c) shows the binned differential distribution measured by the NA62 experiment vs the SM and DM S-scenario contributions. The access events are distributed smoothly over the kinematically allowed range $q^2<(M_K-M_\pi)^2$. So, any smoothly distributed excess events would speak in favour of the S-scenario with a small value of the DM fermion mass parameter $m_\chi$.
\item[(iii)]
For the V-scenario, the allowed range of $m_\chi$ demanded by the Belle II data on $B\to K M_X$ is $m_\chi=0.6\pm 0.2$~GeV. However, the region of smaller $m_\chi$ is likewise not totally excluded. Figure~\ref{Fig:2}(b) provides the theoretical differential distributions for both the SM and the DM V-scenario, and Fig.~\ref{Fig:2}(d) presents the binned differential distribution measured by the NA62 experiment vs the SM and DM V-scenario contributions. In this case, a rather particular DM contribution to the decay $K^+\to\pi^+M_X$ emerges for $m_\chi>0.11$~GeV: the contribution is peaked in the~region $q^2\simeq 0.04-0.07$~GeV$^2$ and may be sufficiently large. So, if higher-statistics data establish excess events peaked at large $q^2$, this behavior will give a strong support to the V-scenario with DM fermion mass $m_\chi\simeq 0.11-0.15$~GeV.
\end{itemize}

\acknowledgments
The research of A.\,B.\ and D.\,M.\ was carried out within the scientific program of the National Center for Physics and Mathematics, Section 5 \emph{Particle Physics and Cosmology}. Stage 2026--2027.

\end{document}